\documentclass[twocolumn,showpacs,pra,aps]{revtex4}
\usepackage{array}
\usepackage{amsmath}
\usepackage{graphicx}

\begin{document}

\title{A scheme for direct observation of entanglement for Gaussian
  continuous variables}

\author{M. S. Kim and Jinhyoung Lee}

\affiliation{School of Mathematics and Physics, Queen's University,
  Belfast BT7 1NN, United Kingdom }

\date{\today}

\begin{abstract}
  We suggest an experimentally realizable scheme to test entanglement of
  a mixed Gaussian continuous variable state.  We find that the
  entanglement condition is simplified for the family of Gaussian states
  which are relevant to experimental realization.  The entanglement
  condition is then shown to be directly related to joint homodyne
  measurements.  We show how robust the proposed test of entanglement is
  against imperfect detection efficiency.
\end{abstract}

\maketitle

In the current development of quantum information processing,
entanglement, which is quantum correlation between particles,
plays a critical role. Recently, the generation, manipulation and
application of continuous-variable entangled states have been
extensively studied \cite{Braunstein,Ralph,Plenio,Leuchs} mainly
for Gaussian states. It is because the Gaussian continuous
variables are extremely similar to qubit systems in theoretical
treatment and they are the continuous-variable states that can be
realized in a laboratory.

A state is said to be entangled when it is not separable.  A Gaussian
continuous-variable state is separable if and only if the partial
transpose of its density matrix is non-negative \cite{Simon}.  This
condition is equivalent to the possibility of imposing a positive
well-defined $P$ function to the state after local unitary operations
\cite{Duan,LeeKim}.  In this paper, we show that the entanglement condition is simplified
for a family of Gaussian states which are more relevant to experimental
realization.  We then suggest a scheme to test entanglement using joint homodyne
measurements.

Einstein, Podolsky and Rosen (EPR) paradox is closely related to
entanglement. There have been discussions on the EPR argument for
continuous variables based on Bell's inequality
\cite{Banaszek,Walmsley}.  On the other hand, Reid and Drummond
\cite{Reid} were concerned with the demonstration of the EPR
paradox itself as distinct from Bell's inequalities.  They derived
a noise level which is violated by the EPR paradox.   Silberhorn
{\em et al.} \cite{Leuchs} proved the generation of an EPR
entangled state by the indirect measure of the sub-vacuum noise in
quadrature correlations. Even though the relation of EPR paradox
to entanglement is still to be ravelled, it has been generally
accepted that there is a subtle difference.  We show that Reid and
Drummond's EPR criterion is only a sufficient condition for
entanglement.

A two-mode Gaussian state is represented by a Gaussian Weyl
characteristic function
\begin{equation}
\label{Wigner}
C({\bf x})=\exp(-\frac{1}{2}{\bf x}{\bf V}{\bf x}^T)
\end{equation}
where ${\bf x}=(q_1,p_1,q_2,p_2)$, ${\bf x}^T$ is the transpose matrix
of ${\bf x}$ and ${\bf V}$ is the quadrature matrix defined as $2\langle
\{x_i, x_j\}\rangle$.  The quadrature variables $q_i$ and $p_i$ for mode
$i(=1,2)$ correspond respectively to quadrature operators
$\hat{q}_i=(\hat{a}_i+\hat{a}_i^\dag)/\sqrt{2}$ and
$\hat{p}_i=i(\hat{a}_i-\hat{a}_i^\dag)/\sqrt{2}$ where $\hat{a}_i$ and
$\hat{a}_i^\dag$ are bosonic operators. The quadrature matrix can in
fact be written using $2\time 2$ block matrices ${\bf L_1}$ and ${\bf
  L_2}$ for local quadrature variables and ${\bf C}$ and its transpose
${\bf C^T}$ representing inter-mode correlation,
\begin{equation}
\label{block} {\bf V}=
\begin{pmatrix}
  {\bf L_1} & {\bf C} \\
  {\bf C^T}&{\bf L_2} \\
\end{pmatrix}.
\end{equation}

{\em Lemma 1.} --- If the block matrices ${\bf L_1}={\bf L_2}$ and ${\bf
  C}$ are diagonal, {\em i.e.}, the quadrature matrix of a Gaussian
continuous-variable state has the following form:
\begin{equation}
{\bf V}_0=
\begin{pmatrix}
  n_1 & 0 & c_1 & 0 \\
  0 & n_2 & 0 & c_2 \\
  c_1 & 0 & n_1 & 0 \\
  0 & c_2 & 0 & n_2
\end{pmatrix}
\label{standard-0}
\end{equation}
where $n_1$ or $n_2$ may be smaller than the vacuum limit 1. The state
is separable if and only if
\begin{equation}
\label{separability-condition}
\delta_1 \delta_2\geq 1
\end{equation}
 where $\delta_i=n_i-|c_i|$ for $i=1,2$.

{\em Proof} --- By local unitary squeezing operations, the matrix
(\ref{standard-0}) is transformed into
\begin{equation}
{\bf V}_1=
\begin{pmatrix}
  n & 0 & c & 0 \\
  0 & n & 0 & c^\prime \\
  c & 0 & n & 0 \\
  0 & c^\prime & 0 & n \\
\end{pmatrix}
\label{standard-1}
\end{equation}
where $n=\sqrt{n_1 n_2}$, $c=c_1\sqrt{n_2/n_1}$ and
$c^\prime=c_2\sqrt{n_1/n_2}$.  The factor $n$ is directly related
to the uncertainty principle to satisfy $n\geq 1$.  For the state
with the quadrature matrix (\ref{standard-1}), Simon's
separability criterion \cite{Simon} reads
\begin{equation}
\label{Simon-condition}
(n^2-c^2)(n^2-c^{\prime 2})\geq 2n^2+2|cc^\prime|-1.
\end{equation}
Define the average and the difference of the correlation factors, $|c|$
and $|c^\prime|$: $c_a=(|c|+|c^\prime|)/2$ and $c_d=(|c|-|c^\prime|)/2$.
Using the new parameters $c_a$ and $c_d$, Simon's criterion
(\ref{Simon-condition}) can be written as
\begin{equation}
\label{Simon-factor}
[(n-c_a)^2-(1+c_d^2)][(n+c_a)^2-(1+c_d^2)]\geq 0
\end{equation}
where $[(n+c_a)^2-(1+c_d^2)]$ is positive unless $n=1$ and
$c=c^\prime=0$ when it becomes zero.  The separability condition is
satisfied if and only if
\begin{equation}
\label{Simon-2}
 (n-c_a)^2-(1+c_d^2)\geq 0 \Leftrightarrow (n-|c|)(n-|c^\prime|)\geq 1.
\end{equation}
With use of the definitions of $n$, $c$ and $c^\prime$ for the
inequality in the right-hand side of the arrow we obtain the
separability condition in Eq.~(\ref{separability-condition}).

One can also use Duan {\em et al.}'s separability criterion \cite{Duan}
even though one has to be careful because all the diagonal elements in
their standard form II have to be larger than 1 for their test of
separability.

Now, recalling the
definition of the quadrature matrix, the separability criterion
(\ref{separability-condition}) can be written as
\begin{eqnarray}
\label{separability-quadrature}
&&(\langle q_1^2\rangle + \langle q_2^2\rangle - 2|\langle q_1q_2\rangle|)(\langle
p_1^2\rangle + \langle p_1^2\rangle - 2|\langle p_1p_2\rangle|) \nonumber \\
&\ge&(\langle q_1^2\rangle +\langle q_2^2\rangle-2|\langle
q_1q_2\rangle|)_0(\langle p_1^2\rangle+\langle
p_1^2\rangle-2|\langle p_1p_2\rangle|)_0 \nonumber \\ &&
\end{eqnarray}
where $\langle X\rangle_0$ denotes mean value of $X$ for the vacuum.
The right-hand side of the inequality is 1, which can be easily seen by
substituting $\bar n=0$ and $s=0$ into (\ref{element-sq-th}).  We have found that the
entanglement of a Gaussian field in the form (\ref{standard-0}) can be
tested by comparing the quadrature correlation of the field with that of the
vacuum as shown in Eq.~(\ref{separability-quadrature}).

What is the relevance of the quadrature matrix given in
(\ref{standard-0}) to our study of entanglement for continuous
variables?  Among many possible ways to produce Gaussian entangled
states, two methods are experimentally more relevant: One is to use a
non-degenerated parametric amplifier (NOPA) to produce a two-mode
squeezed state \cite{Braunstein} and the other is to use a beam splitter
as an entangler \cite{Leuchs,Kim02}.

Let us consider the entanglement of the output field from a beam
splitter when two independent Gaussian fields are incident on it.  The
logical definition of the density operator $\hat{\rho}_s$ for a
single-mode continuous-variable Gaussian state is \cite{Gardiner91}
\begin{equation}
\label{Gaussian-single}
\hat{\rho}_s={\cal N}\exp(-j
\hat{a}^\dag\hat{a}-\frac{1}{2}l\hat{a}^{\dag
  2}-\frac{1}{2}l^*\hat{a}^2)
\end{equation}
where ${\cal N}$ is a normalization factor (throughout the paper the
normalization factor is generally denoted by ${\cal N}$ even though the
detailed forms may differ.).  When $l=0$ and $j=\hbar\omega/kT$, where
$k$ is the Boltzmann constant, the density operator
(\ref{Gaussian-single}) represents a thermal state of temperature $T$.
It is straightforward to show that the general Gaussian state
(\ref{Gaussian-single}) may be transformed into a thermal state by the
unitary single-mode squeezing operator \cite{Loudon},
$\hat{S}(\zeta)=\exp(\frac{1}{2}\zeta^*\hat{a}^2-\frac{1}{2}
\zeta\hat{a}^{\dag 2})$ with the complex squeezing parameter
$\zeta=s\mbox{e}^{i\varphi}$:
\begin{equation}
\label{Gaussian-single-temp}
\hat{S}^\dag(\zeta)\hat{\rho}_s\hat{S}(\zeta)={\cal N}\exp(-2\sqrt{n^2-|m|^2}\hat{a}\hat{a}^\dag)
\end{equation}
for $s=(|l|-j)/4(j+|l|)$ and $\varphi=Arg(l)$.

Consider that the input field described by the operator $\hat{a}_1$ is
superposed on the other input field with operator $\hat{a}_2$ by a
lossless symmetric beam splitter, with amplitude reflection and
transmission coefficients $r$ and $t$. The output-field annihilation
operators are given by $\hat{c}_1=\hat{B}\hat{a}_1\hat{B}^\dag$ and
$\hat{c}_2=\hat{B}\hat{a}_2\hat{B}^\dag$ where the beam splitter
operator is \cite{Campos89}
\begin{equation}
\label{eq:Fock-beamsplitter}
\hat{B}=\exp\left[\frac{\theta}{2}(\hat{a}_1^\dag\hat{a}_2\mbox{e}^{i\phi}-\hat{a}_1\hat{a}_2^\dag\mbox{e}^{-i\phi})\right]
\end{equation}
with the amplitude reflection and transmission coefficients
$t=\cos(\theta/2)$ and $r=\sin(\theta/2)$. The phase difference
$\phi$ between the reflected and transmitted fields can be
adjusted by putting a phase shifter.

When the two input fields are Gaussian states, the output state from a
beam splitter is
\begin{equation}
\label{eq:squeezed}
\hat{B}\hat{S}_1(\zeta_1)\hat{S}_2(\zeta_2)\hat{\rho}^{th}_1\hat{\rho}^{th}_2
\hat{S}^\dag_2(\zeta_2)\hat{S}^\dag_1(\zeta_1)\hat{B}^\dag
\end{equation}
where $\rho^{th}$ is the density operator for a thermal state and
the relation (\ref{Gaussian-single-temp}) has been used. Without
losing generality, we take the input squeezing parameter to be
real while keeping $\phi$ of the beam splitter variable.

It has been found by us \cite{Kim02} that two squeezed states may be
maximally entangled when the beam splitter has $\theta=\pi/4$ and
$\phi=0$ in which case
\begin{eqnarray}
\label{simple-squeezing}
&&\hat{B}\hat{S}_1(s_1)\hat{S}_2(-s_2)\hat{B}^\dag\nonumber \\
&&=\hat{S}_1\left(\frac{s_1-s_2}{2}\right)\hat{S}_2\left(\frac{s_1-s_2}{2}\right)
\hat{S}_{12}\left(\frac{s_1+s_2}{2}\right)
\end{eqnarray}
where
$\hat{S}_{12}(\zeta)=\exp(-\zeta\hat{a}_1\hat{a}_2+\zeta^*\hat{a}_1^\dag\hat{a}_2^\dag)$
is the two-mode squeezing operator \cite{Barnett}. Throughout the paper,
$s_1$ and $s_2$ are assumed to be positive.  The two-mode output field
is represented by the quadrature matrix in the form of
(\ref{standard-0}) with its elements
\begin{eqnarray}
\label{elements}
n_1=\frac{\tilde{n}_1}{2}\mbox{e}^{-2s_1}+\frac{\tilde{n}_2}{2}\mbox{e}^{2s_2}
~&;&~ n_2=\frac{\tilde n_1}{2}\mbox{e}^{2s_1}+\frac{\tilde
n_2}{2}\mbox{e}^{-2s_2}
\nonumber \\
c_1=\frac{\tilde n_1}{2}\mbox{e}^{-2s_1}-\frac{\tilde
n_2}{2}\mbox{e}^{2s_2} ~&;&~ c_2=\frac{\tilde
n_1}{2}\mbox{e}^{2s_1}-\frac{\tilde n_2}{2}\mbox{e}^{-2s_2}
\nonumber \\
&&
\end{eqnarray}
where $\tilde n_i=2\bar n_i+1$ with the mean photon number $\bar
n_i$ of the thermal state.  Substituting these into the
separability criterion (\ref{separability-condition}) we find that
the output field is separable if and only if
\begin{equation}
\label{separable-condition}
(\tilde n_1\mbox{e}^{-2s_1})(\tilde n_2\mbox{e}^{-2s_2})\geq 1.
\end{equation}
$\tilde n_i\mbox{e}^{-2s_i}$ is the quadrature variance of the
input field, which represents the sub-vacuum noise level when it
is smaller than 1.

For the generation of a squeezed state using a NOPA, even though
it is desirable to produce a two-mode squeezed vacuum, it may well
be the case that a two-mode squeezed thermal field is produced or
a two-mode squeezed vacuum is produced but decohered in a thermal
bath.  Let us thus consider an two-mode squeezed thermal field
represented by its density operator:
\begin{equation}
\label{density-sq-thermal}
\hat{\rho}^s_{12}(\zeta)=\hat{S}_{12}(\zeta)\hat{\rho}_1^{th}\hat{\rho}_2^{th}\hat{S}_{12}^\dag(\zeta).
\end{equation}
By the local rotation operator
$\hat{R}_1(\varphi)=\exp(i\varphi\hat{a}_1^\dag\hat{a}_1/2)$, the
two-mode squeezed thermal field can be transformed into
$\hat{\rho}^s_{12}(s)$. A local unitary operation does not change
the nature of entanglement and the local unitary rotation is
easily realized by a phase shifter.  Thus we do not lose
generality by studying how to test the entanglement of
$\hat{\rho}^s_{12}(s)$. We assume that the temperatures of thermal
fields represented by $\hat{\rho}_1^{th}$ and $\hat{\rho}_2^{th}$
are the same: $\tilde n_1=\tilde n_2\equiv\tilde n$. In this case,
the quadrature matrix of the squeezed thermal field
$\hat{\rho}^s_{12}(s)$ has the form (\ref{standard-0}) with matrix
elements:
\begin{eqnarray}
n_1 &=& n_2=\tilde{n}\cosh 2s
\nonumber \\
c_1 &=& -c_2=-\tilde n\sinh 2s. \label{element-sq-th}
\end{eqnarray}

On the other hand, if a squeezed vacuum is produced by a
NOPA but decohered in the thermal environment with the mean number of thermal photons
$\bar n$, the decohered state is still represented by the form
(\ref{standard-0}) but with its elements \cite{LeeKim}
\begin{eqnarray}
n_1 &=& n_2=\mbox{e}^{-\gamma t}\cosh 2s+\tilde n(1-\mbox{e}^{-\gamma t})
\nonumber \\
c_1 &=& -c_2=\mbox{e}^{-\gamma t}\sinh 2s,
\label{element-sq-deco}
\end{eqnarray}
where $\gamma$ is the coupling of the field with the environment.  The
influence from the environment to each mode has been assumed same.

We have seen that most of the two-mode Gaussian states represented by the
quadrature matrices in the form (\ref{standard-0}) are
relevant to current experimental techniques.
The quadrature variables are measured by setting a balanced homodyne
detector \cite{Loudon,Yuen}, which is a well-known device to detect
phase-dependent properties of an optical field, at each mode of the
two-mode field. The operational representation of the balanced homodyne
detector is
\begin{equation}
\label{operation-homo}
\hat{O}_{HD}=\hat{q}\cos\chi-i\hat{p}\sin\chi
\end{equation}
where $\chi$ depends on the local-oscillator phase.  Because the phase
of the local oscillator is not absolute, we must find the
local-oscillator phase which gives the off-diagonal terms of ${\bf
  L_1}$, ${\bf L_2}$ and ${\bf C}$ vanish.  It is straightforward to
measure the off-diagonal terms of ${\bf C}$ by joint homodyne
measurement as $V_{14}=2\langle q_1 p_2\rangle$ and $V_{23}=2\langle q_2
p_1\rangle$.  However, measuring the off-diagonal terms of local
matrices are troublesome as it involves the joint measurement of two
quadrature variables for a single mode.

\begin{figure}[htbp]
  \centering
  \includegraphics[width=0.4\textwidth]{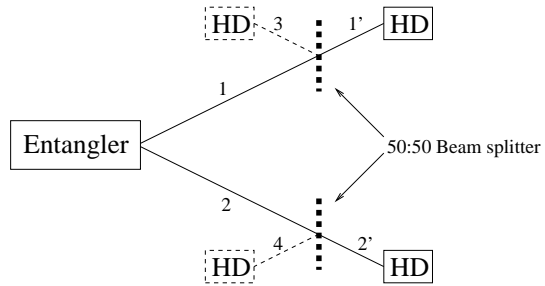}
  \caption{Configuration of the setup to test
    entanglement. The dotted devices measure off-diagonal terms of the
    local quadrature matrices ${\bf L_1}$ and ${\bf L_2}$. The boxed
    device denoted by HD are homodyne detectors. The numbers refer to
    the modes.}
  \label{fig:beam-splitter}
\end{figure}

To measure the off-diagonal elements of ${\bf L_1}$ \cite{heterodyne}, we put a 50:50 beam
splitter which splits the field in mode 1 as  schematically shown in
Fig.~\ref{fig:beam-splitter}. Using the beam splitter
operator (\ref{eq:Fock-beamsplitter}) for the 50:50 beam splitter, we
find that the field for three modes $1^\prime, 2$ and 3 is still
Gaussian and its quadrature matrix is written as
\begin{equation}
\label{block-bs} \frac{1}{2}
\begin{pmatrix}
  {\bf L_1}+\openone &\sqrt{2}{\bf C} &  -{\bf L_1}+\openone \\
  \sqrt{2}{\bf C^T}&{\bf L_2} & -\sqrt{2}{\bf C^T} \\
  -{\bf L_1}+\openone & -\sqrt{2}{\bf C} &  {\bf L_1}+\openone \\
\end{pmatrix},
\end{equation}
where the unit matrix $\openone$ is due to the vacuum injected into the
unused port of the beam splitter. Now the off-diagonal elements of ${\bf
  L_1}$ can be measured by inter-mode correlation between modes
$1^\prime$ and 3: The mean value of the joint measurement $\langle
q_{1^\prime}p_3\rangle=-V_{12}/4$. Similarly other off-diagonal
terms of the local quadrature matrices can be obtained. It is true
that if the entangled field is one of the fields described in this
paper, choosing the phase of the local oscillator to vanish the
off-diagonal terms $V_{14}$ and $V_{23}$, all the local
off-diagonal terms should vanish but to make it sure the above
supplementary measurements can be performed.

In fact, as we have shown, we know how to find all the matrix
elements of the quadrature matrix for a Gaussian field so that it
is possible to test entanglement not only for the fields in the
form (\ref{standard-0}) but also for any Gaussian field if the
detection efficiency is unity. In this case we have to use Simon's
separability criterion \cite{Simon}.

A homodyne detector is composed of two photodetectors. Inefficient
photodetectors introduce noise to each mode and reduces the
quantum correlation between two modes. The detection efficiency
may thus determine the feasibility of the proposed scheme. If the
efficiencies of the photodetectors are same, homodyne measurement
by imperfect detectors is equivalent to homodyne measurement by
perfect detectors following a beam splitter, one input port of
which is fed by the field to be measured and the other by the
vacuum \cite{Leonhardt94}.  The efficiency $\eta$ of the homodyne
measurement determines the transmission coefficient $\sqrt{\eta}$
of the beam splitter.   In fact the fictitious beam splitter
affects the testing field as though it is decohered in the vacuum
reservoir. The detection efficiency, assumed the same for the both
homodyne detectors, effectively changes the quadrature matrix from
${\bf V}$ to ${\bf V}^\prime=\eta {\bf V} + (1-\eta) \openone$.
This is what is measured by imperfect homodyne detectors.  If
${\bf V}$ is in the form ${\bf V}_0$, the variance matrix ${\bf
V}_0^\prime$ takes the same form as ${\bf V}_0$ but with modified
matrix elements $n^\prime_i=\eta n_i + (1-\eta)$ and
$|c^\prime_i|=\eta |c_i|$ for each $i$.

\begin{figure}[htbp]
  \centering
  \includegraphics[width=0.4\textwidth]{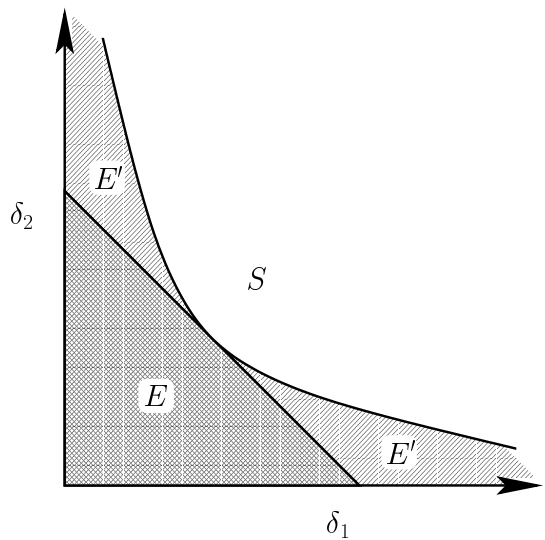}
  \caption{Gaussian states with the symmetric quadrature matrix ${\bf V}_0$ in
    Eq.~(\ref{standard-0}) on the space of $\delta_1=n_1-|c_1|$ and
    $\delta_2=n_2-|c_2|$.  Separable states are denoted by $S$ and
    entangled states by $E$ and $E^\prime$ with the boundary of $\delta_1 \delta_2 =
    1$. Further the two regions $E$ and $E'$ are separated by the line
    $\delta_1+\delta_2=2$. The entanglement imposed on a state
    of the region $E'$ may fail the entanglement test due to inefficient detection.
    However, the entangled state of the region $E$ is always found to be entangled
    regardless of the detection efficiency.}
  \label{fig:ead}
\end{figure}

Consider the effect of the detection efficiency on the
inseparability of the testing fields.  Substituting $n_i^\prime$
and $c_i^\prime$ into Eq.~(\ref{separability-condition}) for
inefficient detection, we find a state to be entangled when
\begin{eqnarray}
  \label{eq:deis}
  (n^\prime-|c_1^\prime|)(n_2^\prime-|c_2^\prime|)
  =\eta^2(\delta_1\delta_2-\delta_1-\delta_2+1)
  \nonumber \\
  +\eta(\delta_1+\delta_2-2)+1<1.
\end{eqnarray}
Rearranging this equation, we can easily find that when the
original testing field is characterized by $\delta_1+\delta_2<2$,
it is always found to be entangled regardless of the detection
efficiency unless the efficiency is zero.  With use of
(\ref{element-sq-th}) we find that {\em a two-mode squeezed
thermal field is extremely robust to the detection efficiency as
an entangled two-mode squeezed state always passes the test of
entanglement regardless of detection efficiency.}
Fig.~{\ref{fig:ead}} presents the sets of Gaussian states on the
space of $\delta_1$ and $\delta_2$ where separable states with
$\delta_1 \delta_2 \ge 1$ are denoted by $S$ and entangled states
with $\delta_1 \delta_2 < 1$ by $E+E'$. {\em All entangled states
in the region $E$ with the condition, $\delta_1 + \delta_2 < 2$,
will violate the inequality (\ref{separability-condition}) of the
quadrature correlation unless the detection efficiency is zero}
while some entangled states in $E'$ fail the test of entanglement.

It is also important to be able to tell a given state is pure. A
degree of purity can be defined as $P=\mbox{Tr} \hat{\rho}^2$.
When $P=1$ the state is pure.   The purity of a two-mode Gaussian
state can be performed using ideal homodyne detectors.  For a
Gaussian state of the quadrature matrix (\ref{standard-0}) the
degree of purity $P= 1 / \sqrt{\det V}_0$. Noting that a vacuum is
pure, the inequality of $P \le 1$ can be written as
\begin{eqnarray}
  \label{eq:pt}
  (\langle q_1^2 \rangle^2 &-& \langle q_1 q_2 \rangle^2 )
  (\langle p_1^2 \rangle^2 - \langle p_1 p_2 \rangle^2 )
  \nonumber \\
  & \ge &
  \left(\langle q_1^2 \rangle^2 - \langle q_1 q_2 \rangle^2
  \right)_0
  \left(\langle p_1^2 \rangle^2 - \langle p_1 p_2 \rangle^2
  \right)_0.
\end{eqnarray}
The equality implies that the testing fields are in the pure state
(with the minimum uncertainty).  For inefficient detectors, once
their efficiencies are known as $\eta$, the purity may be deduced
from the experimental data.

Reid and Drummond derived the inequality for the quantum
correlation between two mode fields along the line with the
Einstein, Podolsky and Rosen argument \cite{Reid,Leuchs}. They
introduced the uncertainty $V_1$ ($V_2$) between the observable
$q_1$ ($p_1$) in one mode and  $q_1^\prime$ ($p_1^\prime$)
inferred from the observation of the other mode. Quantum
correlation may lead the product of the uncertainties to be less
than the vacuum limit, resulting in the inequality of $V_1 V_2 <
1$. In our notation this inequality can be written as
\begin{eqnarray}
  \label{eq:rdi}
  (n_1 - |c_1|)(n_2 - |c_2|) < \frac{n_1 n_2}{(n_1+|c_1|)(n_2+|c_2|)}.
\end{eqnarray}
Note that the right hand side of the inequality is always less
than unity. Thus, the inequality (\ref{eq:rdi}) is sufficient to
satisfy our inseparable condition (see (
\ref{separability-condition})). However, the converse statement
does not hold in general.

We have proposed an experimental scheme to test the entanglement
of a Gaussian field for the first time to the best of our
knowledge.  Our scheme is based on the inseparability criterion.
The present scheme consists of balanced homodyne detectors which
are well-established experimental tools to study quantum optics.
We show that the entanglement imposed in a Gaussian state is
measurable persistently against the detection efficiency if each
mode has the roughly symmetric uncertainty over the phase space.

\acknowledgments

We thank Prof. G. J. Milburn for valuable comments and the UK Engineering and Physical Sciences Research Council for
financial support through GR/R33304.


\begin{references}
\bibitem{heterodyne} This can also be measured using a heterodyne detector.
\end{references}
\end{document}